\documentstyle[11pt]{article}
\def\p#1#2{{\partial#1\over\partial#2}}
\def\go{
\mathrel{\raise.3ex\hbox{$>$}\mkern-14mu\lower0.6ex\hbox{$\sim$}}
}
\def\lo{
\mathrel{\raise.3ex\hbox{$<$}\mkern-14mu\lower0.6ex\hbox{$\sim$}}
}
\begin{document}
\title{RELATIVISTIC ACCRETION}
\author{R. D. Blandford\\
Theoretical Astrophysics, Caltech, Pasadena, CA 91125, USA}
\maketitle
\begin{abstract}
A brief summary of the properties of astrophysical black holes is presented.  
Various modes of accretion are distinguished, corresponding to accretion at rates
from well below to well above the Eddington rate. The importance of mass loss is emphasized
when the accreting gas cannot radiate and it is asserted that
a strong wind is likely to be necessary to carry off mass, angular momentum 
and energy from the accreting gas.   The possible importance of the 
black hole spin in the formation of jets and in dictating the relative importance
of non-thermal emission over thermal radiation is discussed. 
\end{abstract}
\section{Introduction}
The general relativistic theory of black holes was mostly developed in the decade
spanning the discovery of the Kerr (1963) metric to the collation of their properties
in the classic text of Misner, Thorne \& Wheeler (1973), (which is still a necessary
and nearly sufficient reference for most astrophysical
purposes).  Despite 25 years of perspective, it still seems almost
miraculous that equations as complicated as the field equations of general relativity
should produce such an elegant solution and that this should have such magical properties.
After the relativists did their job so well and assured us that black holes {\bf could}
exist, the astrophysicists were left with the
rather messier business of demonstrating that black holes {\bf should} exist
and of calculating (or guessing) their observable characteristics.  At the
stellar level, the first task is a problem in stellar evolution.  Chandrasekhar
(1931, in Cambridge) first showed that there was a maximum mass for a white dwarf;
Oppenheimer \& Volkoff (1939) did the same, in principle, for neutrons stars although
we still do not know its precise value.  However, assuming that it is comfortably
less than $\sim3$~M$_\odot$, we know of about eight, securely-measured, compact
object masses in binary systems with masses well in excess of the Chandrasekhar 
and Oppenheimer-Volkov bounds.  These are black holes, the argument goes;
what else can they be? This is about as strong a degree of ``proof'' as one
typically gets in astronomy and, if one accepts it, the observers have, in turn,
done their job. The {\bf existence} of black holes is no longer an issue.

Turning to massive holes, (with masses of millions to billions
M$_\odot$), it has long been suspected that these lurk in the nuclei of most normal
galaxies, (at least those with luminosities $L\sim L^\ast$), and that they
become active, (and classified as quasars, Seyferts and radio galaxies and so on), 
when fueled at an appropriate rate. Noting again that the most secure evidence comes
from dynamics, ``massive, dark objects'', with masses in the range $\sim2\times10^6$~M$_\odot$
to $\sim3\times10^9$~M$_\odot$, have been located in the nuclei of over 20 nearby 
galaxies.  Among the astrophysical alternatives, that have been discussed
in the past, are ``superstars'', which are 
unstable and would be far too luminous, and clusters of compact objects
which are quite contrived from an evolutionary point of view and have very short dynamical
lifetimes in the best studied examples.  Again, the observers have come through and 
we can conclude, beyond all reasonable doubt, that most large galaxies contain massive
holes in their nuclei.

The most pressing current questions now centre around understanding how accreting
black holes actually behave {\it in situ}. (Evolutionary issues are also
quite important and provide some constraints on this behaviour.)  This type of
research is very difficult to approach deductively from pure physics, because it
involves so many non-elementary process - magnetohydrodynamical turbulence, radiative
transfer, astrophysical particle acceleration and so on.  Therefore, it is prudent to
adopt a more phenomenological approach and to try to formulate astrophysical models involving
techniques that range from order of magnitude estimates to three dynamical 
numerical hydrodynamical simulations, that can meet the burgeoning observational
database in some middle ground.  

In what follows, I will first provide a very quick summary of some salient black hole 
properties and then go on to summarize some properties of Newtonain accretion disks
from a slightly idiosyncratic perspective and emphasizing recent progress that has been made
in understanding the nature of the internal torque.  Next, I will consider some recent
ideas on slow accretion onto black holes.  These ideas are also relevant to the
corresponding problem of fast accretion and here, some possibilities are sketched.  
Another topic, of contemporary interest, which I briefly discuss, 
is the role of black hole spin energy in 
non-thermal emission from, and jet formation by, black holes.
\section{Summary of Black Hole Properties}
Astrophysical black holes, (at least those currently observed), form a two parameter 
family.  They are characterized firstly by their gravitational mass $M$ (as measured
by the orbital period of a distant satellite), which provides a 
scale of length and time through the combinations $GM/c^2\equiv m$ and $GM/c^3$,
respectively. Numerically, 
\begin{equation}
m\equiv1.5\left({M\over M_\odot}\right){\rm km}\equiv5\left({M\over M_\odot}\right)\mu{\rm s}.
\end{equation}
It also furnishes a natural scale for luminosity, the Eddington luminosity
$L_{{\rm Edd}}=4\pi GMc/\kappa_T$, where $\kappa_T$ is the Thomson opacity
and an associated characteristic accretion rate, $\dot M_{{\rm Edd}}\equiv
L_{{\rm Edd}}/c^2$.
(Note that we are not concerned with any quantum mechanical features of black 
holes like Hawking radiation or string entropy.  Astrophysical black holes are
far too large for these effects to be relevant.  They are also far too large
for any electrical charge that they may carry to be of gravitational significance.)

The second parameter is of both geometrical and
dynamical significance.  Traditionally, it is chosen to be the spin angular momentum
per unit mass of the hole (expressed as a length in units of $c$) and denoted by $a$.
This is the quantity that appears in the Boyer-Lindquist form of the metric
({\it e.g.} Misner, Thorne \& Wheeler 1973) and is bounded above 
by $m$. Operationally, it can be measured
by the precession of a distant gyroscope.
However, it is often convenient to use, instead, the angular velocity of the hole, (also
measured in units of $c$ as reciprocal length) and which I shall call $\Omega$.
The relation between these two quantities is given by 
\begin{equation}
\Omega={1-(1-a^2/m^2)^{1/2}\over2a}<{1\over2m}
\end{equation}
$\Omega$ is the angular frequency that an observer at infinity would ascribe to
experimentalists hovering just outside the event horizon (provided that she could 
overcome the strong redshift and see him) ({\it eg} Shapiro
\& Teukolsky 19). These two parameters fully characterize
the geometry of a black hole where the gravitational effect of the
surrounding matter is ignorable and spacetime is asymptotically flat.

Given the Kerr metric, we can compute the orbits of material particles (and photons).
The simplest case is circular orbits in the equatorial plane and this is relevant
to the structure of thin accretion disks.  These have period (measured by a distant
observer)
\begin{equation}
P_K=2\pi(r^{3/2}/m^{1/2}+a)
\end{equation}
when prograde.  These orbits are stable for radii $r>r_{{\rm ms}}$, the marginally 
stable circular orbit.  The corresponding binding energy of this orbit
is denoted $e_{{\rm ms}}$ and increases from 0.06 to 0.42 $c^2$ per unit mass
as $a$ increases from 0 to $m$. It is commonly supposed that gas spirals inward
through the discuss towards the horizon under the action of viscous stress releasing
this binding energy locally as radiation until $r=r_{{\rm ms}}$, when it plunges
quickly into the hole. For this reason, accreting gas is widely thought
to release energy at a rate $\sim10^{20}$~erg g$^{-1}$.

Non-equatorial orbits are more complex.  The most important effect is that their 
orbital angular momentum will precess about that of the hole with a Lense-Thirring
precession frequency, given, to lowest order, by
\begin{equation}
\Omega_{{\rm LT}}=\Omega(r/2m)^{-3}
\end{equation}

One of the more remarkable features of black holes is that they are not truly black.  
A sizeable fraction of their mass energy can be associated with their rotation
and is extractable both in principle and, I assert, in practice.  To make this plausible,
(though not actually {\it prove} anything), allow our experimentalists 
hovering just outside the horizon to be sufficiently thoughtful and to
change the spin angular momentum of the black hole $S$, measured in units of
$G/c^3$, by sending tiny packets of mass energy across the horizon.  This may be in the
form of particles, photons, electromagnetic field etc.  Now, if the angular momentum
is introduced with angular velocity $\omega$, we have 
\begin{equation}
dm=\omega dS=\omega d(am)
\end{equation}
It seems reasonable that if the observers add this angular momentum with 
angular velocity $\omega=\Omega$, then there will be no dissipation. (Just consider
applying a torque to the surface of a spinning disk.) If we spin up a black hole
from rest in this fashion, we can substitute Eq.~2 and integrate differential
Eq.~5 to obtain
\begin{equation}
{m^2\over2}[1+(1-a^2/m^2)^{1/2}]={\rm const}=m_0^2
\end{equation}
where $m_0$ is the initial (or irreducible) mass. The limiting mass to which we
can spin up the hole in this manner is $2^{1/2}m_0$ when $a=m$. 
Equivalently, a little algebra shows that 
\begin{equation}
m_0=m(1+4\Omega^2m^2)^{-1/2}\ge2^{-1/2}m
\end{equation}

This change is reversible.  As Penrose (1969) first showed, there exist negative
energy particle orbits that cross the horizon 
and particles on them {\it decrease} the mass of the 
hole.  
It is then possible to reduce the spin to zero and then return
the mass to $m_0$.  All of this 
becomes more interesting if we use the Kerr metric to compute the area of the horizon
and find that it equates to $A=16\pi m_0^2$.  Therefore, reversible processes are those
that leave the area of the horizon unchanged. If we change the angular momentum
with $\omega\ne\Omega$, the area increases, consistent with its interpretation 
as being proportional to the entropy of the hole. 
(Amusingly, if we define an effective radius $r_0=(A/4\pi)^{1/2}=2m_0$, and define a rotational
speed $\beta=\Omega r_0/c$, we can derive the quasi-Newtonian
relation
$a=r_0^2\Omega$ and the quasi-special relativistic identity $m=m_0(1-\beta^2)^{-1/2}$.)

We can therefore extract an energy, up to the difference
between the gravitational and the irreducible masses,  $m-m_0$, which can be as large as $0.29m$,
from a spinning hole, through Penrose-style processes. However, it turns out that 
this extraction of energy is unlikely to be realized using particles, because it is 
hard to confine them to the requisite negative energy orbits.  The situation is far 
more promising with ordered magnetic field that is supported by external current
({\it eg} Thorne, Price \& MacDonald  1986 and references therein). 
The magnetic field lines can thread the horizon of a spinning black hole.  A very strong
electromotive force will be induced which will make the vacuum into an essentially 
perfect conductor, ({\it eg} through pair-production by $\gamma$-rays), so that the field lines
become equipotentials.  Currents will flow and angular momentum and energy will be exchanged
with the hole.  The relevant angular velocity, $\omega$, is that with which our experimentalist
must move
so that the electric field vanishes.  (If the experimentalist insists upon maintaining
a constant distance from the hole, then this can only be accomplished within a finite range
of radial coordinate.) We can think of this as the angular 
velocity of the magnetic field lines and, in a steady state, it must be constant
along a given field line.  The actual value of this angular velocity depends 
upon the boundary conditions.  Under typical circumstances it is roughly
$\omega\sim0.5\Omega$.  In the frame of an experimentalist hovering 
above the horizon, with an angular velocity $\omega<\Omega$, 
a Poynting flux of electromagnetic energy will be seen
to enter the hole.  However, when we transform into the frame non-rotating
with respect to infinity, we must also include the rate of doing work by the 
electromagnetic torque and we are left with an outwardly directed energy flux
that is conserved along a flux tube. Roughly half of the spin energy of a hole
may be extracted in this manner; the remainder ending up within the horizon 
as an increase in the irreducible mass.
\section{Newtonian Accretion Disks}
First, we review some principles that can be abstracted from the discussion in, for example,
Frank, King \& Raine (1992), Shapiro \& Teukolsky (1983), Pringle (1981), Kato,
Fukue \& Mineshige (1998) and Holt \& Kallman (1998).
Consider a thin disk accretion with angular velocity $\Omega$, 
inflow speed $v \ll \Omega r$, disk mass per unit radius $\mu$ and 
specific angular momentum $\ell$. (Henceforth, we measure all radii
in units of $m$.)
In assuming that the disk is thin, we are implicitly 
supposing that the gas can remain cold by radiating away its internal energy.
Let the torque that the disk interior to radius $r$ exerts 
upon the exterior disk be $G(r)$. The equations of mass and 
angular momentum conservation are then
\begin{equation}
\p\mu t=\p{\mu v}r;\qquad\p{\mu\ell}t=\p{\mu v\ell}r-\p Gr.
\end{equation}  
These equations immediately imply
\begin{equation}
\p Gr={\mu v\ell\over2r};\qquad\p\mu t=2\p{}rr^{1/2}\p Gr
\end{equation}
where we have assumed the Keplerian relation 
$\ell=r^{1/2}$ ({\it cf} Lynden-Bell \& Pringle
1974). 

We can combine Eq.~9 to obtain an energy equation
\begin{equation}
\p{\mu e}t+\p{(\Omega G -\mu v e)}r=G\p\Omega r
\end{equation}
where $e=-\Omega\ell/2$ is the Keplerian binding energy, the sum of
the kinetic and potential energy per unit mass.
(Note the presence of a contribution to the energy flux from the rate at which the
torque $G$, does work on the exterior disk.) The right-hand side represents a radiative loss
of energy. Evaluating it, we find
that the local radiative flux, in a stationary disk,
is three times the rate of local loss of binding 
energy (Lynden-Bell, Thorne in Pringle \& Rees 1972).

Next consider the opposite limiting case when the gas cannot cool and there is no extraneous
source or sink of energy. Adding thermodynamic terms to the energy equation, we obtain
\begin{equation}
\p{\mu(e+u)}t+\p{(\Omega G -\mu v(e+h))}r=G\p\Omega r+\mu T{ds\over dt}
\end{equation}
where $u$ is the vertically-averaged internal energy density, 
$h$ is the enthalpy density, and $s$ is the entropy
density ({\it e.g.} Landau \& Lifshitz 1987).
As there are no sources or sinks of energy, the right-hand side must vanish:
\begin{equation}
\mu T{ds\over dt}=T\left[\p{\mu s}t-\p{\mu vs}r\right]=-G\p\Omega r.
\end{equation}
As the gas has pressure, we must also satisfy the radial equation 
of motion:
\begin{equation}
\p vt-v\p vr+\Omega^2r={1\over r^2}+{1\over\rho}\p Pr.
\end{equation}
\subsection{Magnetic torques}
In order to make further progress, it is necessary to specify the torque, $G$.
A traditional prescription, (Shakura \& Sunyaev 1973), is to suppose
that the shear stress acting in the fluid is directly proportional to the
pressure, with constant of proportionality $\alpha$.  In this case,
\begin{equation}
G=2\pi\alpha r^2\int dzP.
\end{equation}
Traditionally, it has been supposed that $\alpha\sim0.01-0.1$ on the basis of 
unconvincing theoretical and observational arguments.  However, in recent years
a hydromagnetic instability has been rediscovered (Balbus \& Hawley 1998, and references
therein) and it is clear that ionized disks will generate a dynamically
important internal magnetic field on an orbital timescale.  The nature of the
linear instability can be understood by considering a weak, vertical magnetic field line 
threading the disk.  If gas in the midplane is displaced radially outward,
it will drag the magnetic field along with it.  (It is a consequence of electromagnetic
induction in the presence of an excellent conductor, like an ionized accretion disk,
that magnetic field appears to be frozen into the fluid.) As angular momentum is conserved,
the displaced fluid element will lag and stretch the magnetic field lines.  The
magnetic tension associated with the magnetic field 
will have an azimuthal component which will further increase the angular
momentum of the displaced gas and push it further out,
amplifying the instability. (A similar effect is exhibited by a
tethered, artificial satellite.) 

The non-linear 
development of this instability has been investigated numerically and although
many uncertainties remain, it appears that the traditional prescription for
$\alpha$ is not wildly wrong at least as long as the disk is ionized.  (Empirically,
it appears that predominantly neutral disks, as our found in young stellar objects,
for example, exhibit lower values of $\alpha$.)  A major unsolved problem
is the nature of the torque when the accretion rate is large enough for the 
radiation to be trapped by Thomson scattering, so that the disk
fluid becomes radiation-dominated, like the early universe.  Under these
conditions, we expect the short wavelength modes to be damped by radiation drag and
radiative viscosity and the longer wavelength components may escape through buoyancy
(Agol \& Krolik 1998 and references therein).  More numerical simulations, including
radiation transfer, are necessary to help us understand what actually happens.

As we have just emphasized on general grounds, an internal torque in a shearing medium
inevitably leads to dissipation. In the case of MHD torques in an accretion
disk, it has been argued that this happens through a hydromagnetic 
turbulence spectrum which ends up with the 
ions being heated by a magnetic variant on Landau damping called
transit time damping (Quataert \& Gruzinov 1998). 
This is not the only possibility. It is conceivable that 
magnetic reconnection or non-local dissipation in an active, accretion disk
corona may also play a role.
\section{Slow Accretion}
\subsection{ADAF solution}
There has been much attention in recent years to the problem of slow accretion. Observationally,
this is motivated by the discovery that many local galactic nuclei are 
conspicuously under-luminous.  A good example is our Galactic center, where the 
mass supply rate may be as high as $\sim10^{22}$~g s$^{-1}$ and the bolometric
luminosity may be as low as $\sim10^{36}$~erg s$^{-1}$, giving a net efficiency 
of $\sim10^{14}$~erg g$^{-1}$, $\sim10^{-7}c^2$, (and quite unlikely to exceed
$\sim10^{-4}c^2$), a far cry from the naive expectation
of standard disk accretion. As discussed by Narayan \& Yi (1994) and Kato {\it et al} (1998),
and many references therein, one possible resolution of this paradox is that the gas flows
in to the hole as an ``Advection-Dominated Accretion Flow'', or ADAF for short.  In order for
this flow to be established, it is necessary that the gas not be able to cool on the inflow 
timescale.  This, in turn, requires that the viscosity be relatively high and that the hot ions,
which can achieve temperatures as high as $\sim100$~MeV, only heat the electrons by Coulomb
interaction.  (Ultrarelativistic electrons are very efficient radiators.)

The basic idea and assumptions are set out most transparently in 
Narayan \& Yi (1994; cf. also Ichimaru 1977, Abramowicz et al. 1995). In the 
simplest, limiting case, it is assumed 
that there is a stationary, one-dimensional, self-similar flow 
of gas with $\mu\propto r^{1/2}$,
$\Omega\propto r^{-3/2}$, and $v,a\propto r^{-1/2}$, where 
$a=[(\gamma-1)h/\gamma]^{1/2}$ is the isothermal sound speed
and the radial velocity $v<<\Omega r$.
The requirement that $P\propto r^{-5/2}$ transforms the radial equation 
of motion into
\begin{equation}
\Omega^2r^2-{1\over r}+{5a^2\over2}=0.
\end{equation}
Conservation of mass, momentum and energy gives
\begin{equation}
\mu v\equiv\dot M={\rm constant}
\end{equation}
\begin{equation}
\dot Mr^2\Omega-G=F_\ell
\end{equation}
\begin{equation}
G\Omega-\dot M Be=F_E
\end{equation}
where the inwardly directed angular momentum flux, $F_\ell$, and the outwardly directed
energy flux, $F_E$, are constant if there are no sources and sinks of angular momentum
or energy. ($Be=\Omega^2r^2/2-1/r+h$ is the Bernoulli
constant.) Now, the terms on the left-hand side of Eq.~17
scale $\propto r^{1/2}$ and those of  Eq.~18 scale $\propto r^{-1}$.
Therefore, if we require the flow to be self-similar over several decades of radius, 
both constants must nearly vanish.  In the limit,
$F_\ell =F_E=0$.  

Combining equations, we solve for the sound speed $a$ and
the Bernoulli constant.
\begin{equation}
\label{heq}
a^2=\left[{3(\gamma-1)\over5-3\gamma}\right]\Omega^2r^2={6 (\gamma-1)\over(9\gamma-5)r}
\end{equation}
\begin{equation}
\label{bern}
Be=\Omega^2 r^2.
\end{equation}
The elementary ADAF solution is then completed by defining 
an $\alpha$ viscosity parameter through,  $G=\dot Mr^2\Omega=\alpha\mu ra^2$,
which then implies $v={\alpha a^2 /\Omega r }$, 
assuming that $\alpha \ll (5/3-\gamma)^{1/2}$. 

There are concerns with this solution, as identified by Narayan \& Yi (1995).
The most important of these is the worry that the accreting 
gas may not be bound to the black hole. This can be
demonstrated by observing that the Bernoulli constant, $Be$ 
is generically positive due to the viscous
transport of energy.  This means that an 
element of gas has enough internal energy, (taking into account the capacity
to perform $PdV$ work), to escape freely to infinity). In the particular
case when the specific heat ratio is $\gamma\sim5/3$, as it will be
if only the ions are effectively heated, note that the 
the self-similar solution is nearly 
non-rotating.  A lot of angular momentum and orbital kinetic energy must be lost at 
some outer radius, where the ADAF solution first becomes valid.  (This is
called the transition radius.) As the gas cannot cool here, by assumption,
there seems nowhere for the energy to go except in driving gas away.
Another precarious part of the ADAF solution is found close to the 
rotation axis. It is proposed that when the viscous torque is 
relatively large, that the flow extend all the 
way to the polar axis (Narayan \& Yi 1995).  This removes one exposed surface,
but it does so at the expense of creating a stationary column of gas, which 
cannot be supported at its base. It is 
unlikely to persist. 
\subsection{ADIOS solution}
For these reasons, Blandford \& Begelman (1998) have proposed a variant upon 
the ADAF solution called an ``Advection-Dominated Inflow Outflow Solution''.
Here the key notion is that the excess energy and angular momentum is removed by a wind
at all radii.  Again it is simplest to assume self-similarity.  We follow the 
Narayan \& Yi (1984) solution, but supplement it by allowing the mass accretion rate to 
vary with radius.
\begin{equation}
\label{wdotm}
\dot M\propto r^p;\qquad0\le p<1 .
\end{equation}
The mass that is lost from the inflow escapes as a wind.
If we adopt self-similar scalings, and use the above definitions of the flow of
angular momentum and energy, we can write,
\begin{equation}
\label{wfl}
F_\ell=(\dot Mr^2\Omega-G)=\lambda\dot Mr^{1/2};\qquad\lambda>0.
\end{equation}
and
\begin{equation}
\label{wfe}
F_E=G\Omega-\dot M\left({1\over2}\Omega^2r^2-
{1\over r}+{5a^2\over2}\right)={\epsilon\dot M\over r};\qquad\epsilon>0.
\end{equation}
where $\lambda,\epsilon$, like $p$ are constants that can be fixed
Equivalently, for the specific
angular momentum and energy carried off by the wind, we have
\begin{equation}
\label{wdfl}
{dF_\ell\over d\dot M}={\lambda(p+1/2)r^{1/2}\over p};\qquad
{dF_E\over d\dot M}={\epsilon(p-1)\over pr}
\end{equation}

With these modifications, the radial equation of motion becomes
\begin{equation}
\label{wom}
\Omega^2r^2-{1\over r}+(5/2-p)a^2=0.
\end{equation}
Similarly, the Bernoulli constant becomes
\begin{equation}
\label{wbe}
Be={\Omega^2r^2\over2}-{1\over r}+{5a^2\over2}
=pa^2-{1\over2}\Omega^2r^2
\end{equation}
and it can now have either sign. (A limit must be taken to recover Eq.~20.)
Combining these equations, we obtain
\[
\Omega r^{3/2}={(5-2p)\lambda\over 15-2p }
\]
\begin{equation}
\label{womeg}
+\quad{[(5-2p)^2\lambda^2+(15-2p)(10\epsilon+4p-4\epsilon p)]^{1/2}
\over 15-2p}
\end{equation}
It is a matter of algebra to complete the solution and determine 
how the character of the solutions depends upon our  
three independent, adjustable parameters,
$p,\lambda,\epsilon$.

Let us consider some special cases.
\begin{enumerate}
\item $p=\lambda=\epsilon=0$. There is no wind and the system reduces to the non-rotating
Bondi solution.
\item $p=\lambda=0$, $\epsilon=3(1-f)/2$. This corresponds to 
flow with no wind but with radiative loss, which carries away energy 
but not angular momentum. The parameter $f$, (Narayan \&Yi 1994), 
is defined by the relation $\dot MTdS/dr=fGd\Omega/dr$.
\item $p=0$, $\lambda=1$, $\epsilon=1/2$. This describes a magnetically-dominated 
wind with mass flow conserved in the disk. All of the angular 
momentum and energy is carried off by a wind with $dF_E/dF_\ell=
\Omega$ (cf. Blandford \& Payne 1982, K\"onigl 1991). 
There is no dissipation in the disk, which is cold and thin. 
\item $\lambda=2p[(10\epsilon+4p-4\epsilon p)
/(2p+1)(4p^2+8p+15)]^{1/2}$. This corresponds to 
a gas dynamical wind where $dF_\ell/d\dot M\equiv\ell_W=r^2\Omega\equiv\ell$. 
The wind carries off its own angular
momentum at the point of launching and does not exert any reaction torque on the remaining
gas in the disk. Any magnetic coupling to the
disk implies $\ell_W>\ell$.
\item $ra^2=r^3\Omega^2/2p=1/(p+5/2)$. This corresponds to a 
marginally bound flow with vanishing Bernoulli constant. In practice, it is expected
that $Be<0$. In the limiting case, a single proton at the event horizon
can, altruistically, sacrifice itself to allow up to a thousand of its fellow
protons to escape to freedom from $\sim1000m$.
\end{enumerate}

What this exercise demonstrates is that gas can accrete slowly onto a black hole
without radiating, provided it uses loses enough mass, energy and angular momentum
and that the rate of mass accretion by the black hole may be very much less 
than the mass supply rate.  (This has implications for the rate 
of balck hole growth due to accretion in the early universe.) In order to 
go beyond this, we must introduce some additional physics into our discussion
of the disk and the wind.
\section{Fast Accretion}
The solution, that I have just described, is appropriate when the accretion rate is
slow enough that the gas cannot cool radiatively. How slow this must be depends upon
microphysical details that are still uncertain. However, it appears that the underlying 
physical principles are still appropriate in the opposite limiting case of fast 
accretion.  In this limit, as the accretion rate is high the density is sufficiently 
large to allow the gas to come into local thermal equilibrium and to emit radiation
so that the photons dominate the gas pressure.  However, the density is also large enough
for the gas to become optically thick to Thomson scattering and for the radiation 
to be trapped.  Under such circumstances, photons will random walk relative to the 
gas with a characteristic speed $\sim c/\tau_T$, where $\tau_T$ is the optical
depth to Thomson scattering.  If the density is so large that this speed is less than
the bulk speed of the gas, then the photons are essentially trapped and, once again,
the gas is prevented from radiative cooling. Typically, this occurs if $\dot M>
\dot M_{{\rm Edd}}$. 

It is therefore possible to develop a model of accretion onto a black hole in the limit
when $\dot M>>\dot M_{{\rm Edd}}$ (Begelman \& Blandford in preparation).  
We treat disk accretion in much the same way as we treat it in the ion-dominated
case, with the unimportant difference that the effective specific heat ratio
(not the true one) is $\gamma=4/3$. In order to define a vertical structure
for the disk, we have to make some assumption about the angular velocity
distribution. One possibility is that the angular velocity is
constant on cylinders. This is equivalent to assuming that the
equation of state is barytropic.  A much better assumption, and one that
has some physical support, is that the disk is marginally unstable to convective 
overturn ({\it eg} Begelman \& Meier 1982).  This implies that the entropy is constant on surfaces of constant
angular momentum - the ``gyrentropic hypothesis'' ({\it cf} Abramowicz \& Paczy\'nski
1982, Blandford, Jaroszy\'nski \& Kumar 1985). This allows entropy and 
angular momentum to be freely transported along these surfaces; transport
perpendicular to these surfaces requires additional (and presumably magnetic)
viscous stress.

The attached wind is also radiation-dominated
and it is possible to find self-similar solutions that describe a wind
that carries off the mass, angular momentum and energy released from the surface 
of the disk. Eventually, this outflow will become tenuous enough to allow the trapped 
radiation to escape.  There may be a third region, where the flow is optically thin,
and where the gas may start to recombine so that it can be accelerated by line
radiation pressure.

Super-critical accretion flows, with $\dot M>>\dot M_{{\rm Edd}}$, almost surely
occurs naturally in both Galactic sources like SS433 and GRS 1915+112 and in extragalactic
sources like the radio-quiet quasars and the broad absorption line quasars.  
In both circumstances, it appears that the rate of mass outflow exceeds the Eddington
rate by a large factor.  Presumably, the same is true of the rate of mass supply. 
\section{The Importance of Spin}
As emphasized above, there are two potential power sources, the binding energy released
by accreting gas and the spin energy of the central black hole.  It is natural to associate
the former with ``thermal'' emission and the latter with ``non-thermal'' emission and this
separation has provided the basis for a variety of ``unified'', (and ``grand unified'') 
models of AGN over the past twenty years. It is apparent that the electromagnetic
extraction of energy and angular momentum from the black hole can occur in principle.
A question of recent interest is ``How much does this occur in practice?''.

Clearly, there are two requirements beyond the presence 
of the black hole, spin and magnetic flux, and on these we can only
speculate.  It is widely assumed that freshly formed holes, and those that
have recently undergone major mergers, spin rapidly in the sense that $\Omega m\sim0.2-0.5$.
However, subsequent addition of angular momentum (eg through rapid episodes
of accretion or minor mergers) may be stochastic, as opposed to the mass which
increases secularly.  This can lead to a spin down.  Alternative a strong
dynamical interaction without a surrounding warped disk (as proposed
by Natarajan \& Pringle, 1998, preprint) may lead to a very rapid
de-spinning of the hole without the creation of much non-thermal energy.)
It has also been suggested (Ghosh \& Abramowicz 1998,
Livio, Ogilvie \& Pringle 1998, {\it cf} Blandford
\& Znajek 1977) that the total electromagnetic power that derives
from the hole is very small compared with that which derives
from the disk.  The basis of this argument is that the strength of the 
magnetic field that threads the hole is likely to be no larger than
that threading the disk and the area of the disk is much larger
than that of the hole. This, indeed, may be the case in the majority of active 
nuclei (in particular radio quiet-quasars and Seyfert galaxies)
which are not non-thermal objects.

However, it is not guaranteed that these limits always
apply. For example, the strength of the magnetic field interior to the
disk is really only limited by the Reynolds' stress of the orbiting gas
just as is supposed to occur at the Alfv\'en surface surrounding
an accreting, magnetized neutron star.  Alternatively, the magnetic
field associated with the disk may be predominately closed with little radial
component (as it is being strongly sheared) so that it does not extract
much energy, but can provide pressure. Under either of these circumstances 
the non-thermal power extracted from the hole can be dominant.  What actually
occurs depends upon issues of stability and supersonic reconnection.

There is a strong observational incentive to consider these processes.  It
has become clear that some sources are spectacularly non-thermal. Bulk
Lorentz factors in excess of 10 are required to explain some superluminal motion
and, perhaps, much larger relativistic speeds are indicated by the intraday 
variable sources.  The $\gamma$-ray jets discovered by the EGRET instrument
on Compton Gamma Ray Observatory can be prodigiously energetic and 
in some cases, seem to transport far more energy, even allowing for
relativistic beaming, than is observed in the remainder of the 
electromagnetic spectrum (Hartmann {\it et al} 1996).  
The rapidly variable X-rays produced by the Galactic
superluminal sources are far too energetic to be the result of black body-emission
from the surface of an accretion disk.  To this reviewer, at least, it is 
unlikely that this power can derive solely from an active disk
corona. There is ample spin energy associated with the hole
to account for the observations and an environment
where thermalization will be very difficult. The assocation of the jet power
and the high energy emission with the black hole is surely as suggestive 
as, historically, was the association of the Crab Nebula with 
PSR 0531+21.

There is another interesting possibility (Blandford \& Spruit 1998,
in preparation, {\it cf} also Livio, these proceedings).  This is that
magnetic field attached to the inner disk may also connect to the black hole.
Now if $\Omega>0.093/m$, the hole rotates faster than the gas in the marginally
stable circular orbit and even faster than all the gas beyond this.  Therefore,
unless the hole is very slowly rotating, a magnetic connection will transport angular
momentum radially outward.  As the hole has a much higher effective resistivity than
the disk, we can regard the field lines as being effectively transported
by the disk.  Therefore, they will only do mechanical work on the disk with no direct
dissipation. If this interaction is strong enough the increase in the angular momentum
of the gas can be enough to reverse the accretion flow, driving some gas radially outward
while a fraction falls inward.  This can happen in a quasi-cyclical manner and it 
is tempting to associate some of the quasi-periodic behavior observed in Galactic
black hole binaries, notably GRS 1915+112 with just this sort of process.
\section{Conclusion}
I hope that this somewhat cursory description of recent developments is sufficient
to persuade the reader that the study of black holes both in our Galaxy,
in the nuclei of nearby galaxies and in distant quasars, is on the ascendant.
Measurements of mass and (possibly) spin are helping make astrophysical
arguments quantitative.  Observations of quasi-periodicity,
notably by RXTE (Greiner, Morgan \& Remillard 1996), are strongly suggestive
of non-linear dynamical processes in the curved spacetime close to the black hole.
Direct measurement of iron lines and their profiles explore the surfaces 
of inner accretion disks. Jets, are now observed to be commonplace in 
accreting systems and, in the case of black hole systems must derive
from close to the hole (as close as $\sim60m$ inthe case of M87, Junor 
\& Biretta 1995).

The best is yet to come.  There is a suite of X-ray telescopes planned for 
launch in the coming years, AXAF, XMM and ASTRO E.  There are ambitious
plans for superior instruments, like GLAST and CONSTELLATION-X to be launched
during the coming decade and for space-based VLBI to be developed in earnest.
On an even longer timescale, the space-based gravitational radiation
detection, LISA has the projected capability to detect
merging black holes from cosmological distances and to provide direct
quantitative tests of strong field general relativity for the first time.
On the theoretical front, the numerical capability to perform large
scale, three dimensional, hydromagnetic simulations is already here
and the incorporation of radiative transfer and credible dissipative
processes is a not so distant hope.  This is a good time for a student
to start research in black hole astrophysics.

\section*{Acknowledgments}
I am indebted to Mitch Begelman and Henk Spruit 
for recent collaboration and to Charles Gammie, Andy Fabian, 
John Hawley, Mario Livio, Ramesh Narayan,
Eve Ostriker, Jim Pringle and Martin Rees for stimulating discussions.  Support 
under NSF contract AST 95-29170 and NASA contract 5-2837 and the Sloan Foundation
is gratefully acknowledged. I thank John Bahcall and the Institute for Advanced Study 
for hospitality during the completion of this paper.
\section*{References}
\parindent=0pt
\def\reference{\par}
\reference Abramowicz, M. A. \& Paczy\'nski, B. 1982 ApJ 253 897
\reference Agol, E. \& Krolik, J. 1998 ApJ in press
\reference Balbus, S. A. \& Hawley, J. F. 1998 RMP 70 1
\reference Begelman, M. C. \& Meier, D. L. 1982 ApJ 253 873
\reference Blandford, R. D. \& Begelman, M. C. 1998 MNRAS in press
\reference Blandford, R. D. Jaroszy\'nski, M. \& Kumar, S. 1985 MNRAS 225 667
\reference Blandford, R. D. \& Payne, D. G. 1982 MNRAS 199 883
\reference Blandford, R. D. \& Znajek, R. L. 1977 MNRAS 179 433
\reference Chandrasekhar, S. 1931 ApJ 74 81
\reference Frank, J., King, A. \& Raine, D. 1992 Accretion Power in Astrophysics
Cambridge: Cambridge University Press
\reference Ghosh, P. \& Abramowicz, M. A. 1997 MNRAS 292 887
\reference Greiner, J., Morgan, E. H. \& Remillard, R. A. 1996 ApJ 473 L107
\reference Hartmann, R. C. et al. 1996 ApJ 461 698

\reference Holt, S. S. \& Kallman, T. R. ed. 1998 Accretion Processes in Astrophysical
Systems: Some Like it Hot New York: AIP
\reference Junor, W. \& Biretta, J. A. 1995 AJ 109 500
\reference Kato, S., Fukue, J. \& Mineshige, S. 1998 Black Hole Accretion Disks
Kyoto: Kyoto University Press
\reference Kerr, R. P. 1963 PRL 11 237
\reference Landau, L. D. \& Lifshitz, E. M. 1987 Fluid Mechanics Oxford: Butterworth-Heinemann
\reference Lynden-Bell, D. \& Pringle, J. E. 1974 MNRAS 168 603
\reference Misner, C. W. Thorne, K. S. \& Wheeler, J. A. 1973 Gravitation San Francisco: Freeman
\reference Narayan, R. \& Yi, I. 1994 ApJ 428 L13
\reference Narayan, R. \& Yi, I. 1995 ApJ 444 231
\reference Oppenheimer, J. R. \& Volkoff, G. M. 1939 Phys Rev 55 374
\reference Penrose, R. 1969 Riv Nuovo Cim 1 252
\reference Pringle, J. E. 1991 ARAA 19 137
\reference Pringle, J. E. \& Rees, M. J. 1972 A \& A 21 1
\reference Quataert, E. \& Gruzinov, A. 1998 ApJ in press
\reference Shakura, N. I. \& Sunyaev, R. A. 1973 A \& A 24 337
\reference Shapiro, S. L. \& Teukolsky, S. A. 1983 Black Holes, White Dwarfs and Neutron
Stars New York: Wiley
\reference Thorne, K. S., Price, R. M. \& MacDonald 1986 Black Holes: The Membrane
Paradigm New Haven: Yale University Press
\end{document}